\documentclass[12pt]{article}
\newcommand{\text}{\rm}

\begin{document}

\title{{\bf An Algebraic Criterion for the Ultraviolet Finiteness of Quantum Field
Theories}}
\author{V.E.R. Lemes$^{\,{\rm (a)}}$, M.S. Sarandy$^{\,{\rm (b)}}$, S.P. Sorella$^{\,%
{\rm (a)}},\;$ \and O.S. Ventura$^{\,{\rm (a,c)}}\;$and L.C.Q. Vilar$^{\,%
{\rm (a)}}$ {\bf \ }\vspace{2mm} \\
%EndAName
{\small {$^{{\rm {(a)}}}${\it \ UERJ, Universidade do Estado do Rio de
Janeiro}}} \\
{\small {{\it \ Rua S\~{a}o Francisco Xavier 524, 20550-013 Maracan\~{a},
Rio de Janeiro, Brazil.}}}\vspace{1mm}\\
{\small {$^{{\rm {(b)}}}\,${\it CBPF, Centro Brasileiro de Pesquisas
F\'{\i}sicas}}}\\
{\small {\it \ Rua Xavier Sigaud 150, 22290-180 Urca, Rio de Janeiro, Brazil.%
}}\vspace{1mm}\\
{\small {$^{{\rm {(c)}}}\,${\it Col\'{e}gio Rio de Janeiro, Rua M. Rubens
Vaz 392,}}}\\
{\small {\it \ 22470-070 Gavea, Rio de Janeiro, Brazil.}}\vspace{1mm}}
\maketitle

\begin{abstract}
An algebraic criterion for the vanishing of the beta function for
renormalizable quantum field theories is presented. Use is made of the
descent equations following from the Wess-Zumino consistency condition. In
some cases, these equations relate the fully quantized action to a local
gauge invariant polynomial. The vanishing of the anomalous dimension of this
polynomial enables us to establish a nonrenormalization theorem for the beta
function $\beta _g$, stating that if the one-loop order contribution
vanishes, then $\beta _g$ will vanish to all orders of perturbation theory.
As a by-product, the special case in which $\beta _g$ is only of one-loop
order, without further corrections, is also covered. The examples of the $%
N=2,4$ supersymmetric Yang-Mills theories are worked out in detail.
\end{abstract}

\vfill\newpage\ \makeatother

\renewcommand{\theequation}{\thesection.\arabic{equation}}

\section{Introduction}

The search for ultraviolet finite renormalizable models has always been one
of the most attractive and relevant aspect of quantum field theory. The
requirement of a softer ultraviolet behavior has motivated the construction
of many models, including the Yang-Mills supersymmetric theories (SYM), the
supergravities as well as the superstrings.

The ultraviolet finiteness is understood here as the vanishing, to all
orders in the perturbative loop expansion, of the beta functions of the
theory. This means that the dependence from the renormalization scale can be
fully accounted by the unphysical anomalous dimensions of the field
amplitudes which are, in general, nonvanishing.

So far, many ultraviolet finite theories have been found in different
space-time dimensions. For instance, the Wess-Zumino-Witten models \cite{wzw}
and the $N=\left( 4,0\right) $ supersymmetric $\sigma $-model \cite{sig} are
examples of two-dimensional theories which turn out to be conformal and
superconformal, respectively.

In three space-time dimensions, the so called topologically massive
Yang-Mills theory, obtained by adding the Chern-Simons action to the
Yang-Mills term, is one of the most celebrated example of a fully\footnote{%
In this case the anomalous dimensions of the fields vanish as well.} finite
theory \cite{djt,pis,vg} with applications in QCD at nonzero temperature.
Also, the pure Chern-Simons theory is known to have vanishing beta function
and field anomalous dimensions to all orders of perturbation theory \cite
{witten,csl,bc,cssusy,ghe}. Its topological nature has allowed to use
perturbative techniques to evaluate topological invariants of knots theory 
\cite{topinv}. The beta function corresponding to the Chern-Simons
coefficient vanishes in the presence of matter as well \cite
{gs,csm2,csm1,ymcs1}. More generally, in the abelian case this coefficient
is known to be strongly constrained by the Coleman-Hill theorem \cite
{coleman}, implying that it can receive at most one-loop finite corrections.
We remark that the one-loop induced Chern-Simons coefficient has an
important physical meaning, identifying indeed the transverse conductivity.
In addition, as shown in \cite{matsuyama}, this coefficient turns out to be
quantized by a topological argument. It is worth mentioning here that,
recently, the Coleman-Hill theorem has been extended to the nonabelian case 
\cite{das}.

Turning now to four dimensions, the supersymmetric gauge theories certainly
display a unique ultraviolet behavior, leading in some cases to finite
renormalizable field theories. This is the case of $N=4$ SYM, which provided
the first example of a four-dimensional superconformal gauge theory \cite
{n4l,n4w}. Concerning the $N=2$ SYM, although it\ is not ultraviolet finite,
its beta function obeys a remarkable nonrenormalization theorem, stating
that it receives at most one-loop contributions \cite{n2l,n2gri}. In the
case of $N=1$ SYM\ a set of necessary and sufficient conditions for the
vanishing of the beta function to all orders of perturbation theory has been
established \cite{n1}, making possible to classify the $N=1$ finite SYM
theories.

Examples of higher dimensional finite field theories are provided by the $BF$
models \cite{bf}, which belong to the class of the Schwarz type topological
field theories \cite{schwarz}.

The ultraviolet finiteness of the above mentioned theories has been checked
first by explicit loop calculations \cite{djt,pis,vg,csl,gs,n4l,n2l}, and
later on has been proven, to all orders of perturbation theory, by using a
suitable set of Ward identities characterizing the symmetry content of each
model.

For instance, in the case of the $(4,0)$ two-dimensional $\sigma $ model the
use of the BRST\ technique has allowed for a regularization independent
proof of the absence of the superconformal anomaly \cite{bp2}. A BRST
approach has also been employed in the case of the Wess-Zumino-Witten models 
\cite{bp1} and of the four-dimensional $N=4$ SYM \cite{white,n4sor}.

Concerning the $N=2$ SYM, the proof of the nonrenormalization theorem of its
beta function given in ref.\cite{n2sor} is based on a key relationship
between the whole action of $N=2\;$and a local gauge invariant polynomial
which turns out to have vanishing anomalous dimension. A different proof of
this theorem is available also within the context of the harmonic superspace 
\cite{buchbinder}.

A detailed analysis of the quantum properties of the supercurrent multiplet
is at the basis of the finiteness conditions for $N=1$ SYM theories \cite{n1}%
.

The vanishing of the beta function for the topological field theories can be
proven in a rather general way by making use of an additional nonanomalous
symmetry, called vector supersymmetry, present in both Schwarz and Witten's
type theories \cite{Delduc,ms}. The existence of this further symmetry
relies on the fact that the energy-momentum tensor of the topological
theories can be cast in the form of a pure BRST variation. We also underline
that the trace of the energy momentum tensor, whose integrated quantum
extension is directly related to the beta function, can be characterized by
a set of Ward identities based on a local formulation of the dilation
invariance \cite{dilat,bc,csm1,ymcs1,barnich}. This approach has been
successfully applied to pure Chern-Simons \cite{bc} and to topologically
massive Yang-Mills \cite{csm1,ymcs1}. In this latter case a different proof
of the finiteness has been given in ref.\cite{ymcs2}, using a cohomological
argument for a generalized class of Yang-Mills theories.

Besides the use of Ward identities, the reduction of couplings introduced by
Oehme and Zimmermann \cite{oehme} provides a very powerful and original
method in order to reduce the number of independent coupling constants
present in a given model. The requirement that the reduced theory has fewer
independent couplings leads to a nontrivial set of reduction equations,
relating the various beta functions. Although some of the relationships
between the couplings can be associated to the existence of symmetries, one
has to observe that the solutions of the reduction equations do not always
seem to correspond to any known invariance \cite{oehme}.

The aim of this work is to present a purely algebraic criterion, of general
applicability, for the ultraviolet finiteness. The approach relies on the
BRST cohomology \cite{bbh} and exploits the set of descent equations
following from the Wess-Zumino consistency condition. It turns out indeed
that, in some cases, these equations allow to establish a one to one
correspondence between the quantized action of a given model and a local
field polynomial, belonging to the cohomology of the BRST operator in the
lowest level of the descent equations. As a consequence, the beta function
of the theory can be proven to be related to the anomalous dimension of this
polynomial. The absence of this anomalous dimension entails therefore a
nonrenormalization theorem for the beta function. This theorem states that
if the beta function vanishes at one-loop order, it will vanish to all
orders of perturbation theory, implying the ultraviolet finiteness of the
model. As a by-product, it will be possible to cover also the case in which
the beta function happens to be only of one-loop order, without any further
corrections. The advantage of this approach is that the anomalous dimension
of the field polynomial to which the quantized action is related, is easier
to control than the proper beta function thanks to the existence of
additional Ward identities as for instance the ghost equation \cite{ghe},
always present in the Yang-Mills type theories in the Landau gauge.

The paper is organized as follows. In Sect.2 the general assumptions needed
for the finiteness theorem are discussed. Sect.3 is devoted to the proof of
the theorem, including the analysis of the absence of higher order
corrections for the beta function. In Sect.4 several examples will be worked
out. These include the case of the Chern-Simons coupled to matter, the $N=2$
and the $N=4$ SYM theories in four dimensions. Finally, in Sect.5 we
summarize our main results, presenting the conclusion.

\section{The general set up}

\subsection{ Classical aspects}

Let us start by fixing the notations and by specifying the classical and the
quantum assumptions about the structure of the models which will be
considered throughout. We shall work in a flat $\,D\,$-dimensional euclidean
space-time equipped with a set of fields generically denoted by $\{\Phi ^i\}$%
, $i$ labelling the different kinds of fields needed to properly quantize
the model, {\it i.e.} gauge fields, matter fields, ghosts, ghosts for
ghosts, etc. According to the Batalin-Vilkovisky quantization procedure \cite
{batalin}, for each field $\Phi ^i$ with ghost number ${\cal N}_{\Phi ^i}$
and dimension $d_{\Phi ^i}$, one introduces a corresponding antifield $\Phi
^{i*}$ with ghost number $-(1+{\cal N}_{\Phi ^i})$ and dimension $(D-d_{\Phi
^i})$.

We shall start thus with a classical fully quantized action $\Sigma (\Phi
^i,\Phi ^{i*})$ which will be considered to be massless and, for simplicity,
to have a unique coupling constant $g$. The action $\Sigma (\Phi ^i,\Phi
^{i*})$ is power-counting renormalizable and obeys the classical
Slavnov-Taylor identity

\begin{equation}
{\cal S}(\Sigma )=\int d^Dx\,\frac{\delta \Sigma }{\delta \Phi ^i}\frac{%
\delta \Sigma }{\delta \Phi ^{i*}}=0\,,  \label{st}
\end{equation}
which leads to the nilpotent linearized operator $B_\Sigma $ 
\begin{equation}
B_\Sigma =\int d^Dx\,\left( \frac{\delta \Sigma }{\delta \Phi ^i}\frac
\delta {\delta \Phi ^{i*}}+\frac{\delta \Sigma }{\delta \Phi ^{i*}}\frac
\delta {\delta \Phi ^i}\right) \;,\;\;\;\;\;\;B_\Sigma B_\Sigma =0\;.
\label{stl}
\end{equation}
Concerning the dependence from the coupling constant $g$, we shall make use
of the following parametrization

\begin{equation}
\Sigma =\frac 1{g^2}\int d^Dx{\cal L}_{{\rm inv}}\;+\Sigma _{{\rm gf}%
}\;+\Sigma _{{\rm \Phi }^{*}}\;,  \label{param}
\end{equation}
where ${\cal L}_{{\rm inv}}$ is the classical invariant lagrangian
identified as the part of $\Sigma $ which is independent from the
antifields, the ghosts and the Lagrange multipliers, entering respectively
the gauge-fixing term $\Sigma _{{\rm gf}}$ and the antifield action $\Sigma
_{{\rm \Phi }^{*}}$. As is well known, with this parametrization a $L$-loop
Feynman diagram behaves as $g^{2(L-1)}$.

Differentiating now the Slavnov-Taylor identity $\left( \ref{st}\right) $
with respect to the coupling constant $g$, we obtain the equation

\begin{equation}
B_\Sigma \frac{\partial \Sigma }{\partial g}=0\;,  \label{cond}
\end{equation}
showing that $\partial \Sigma /\partial g$ is an invariant cocycle.
Actually, according to the requirement that $g$ is a physical parameter of
the theory, the cocycle $\partial \Sigma /\partial g$ turns out to be
nontrivial\footnote{%
It can be proven \cite{book} that physical quantities, such as the Green's
functions of gauge invariant operators, are independent from a parameter $%
\alpha $ for which $\partial \Sigma /\partial \alpha $ is trivial, {\it i.e. 
}$\partial _\alpha \Sigma =B_\Sigma \Xi $ for some local polynomial $\Xi $.
Such a parameter is called a gauge parameter.}, identifying therefore the
cohomology of the operator $B_\Sigma $ in the sector of the integrated local
polynomials with ghost number zero and dimension $D.$

Owing to the parametrization $\left( \ref{param}\right) ,$ it follows that

\begin{equation}
\frac{\partial \Sigma }{\partial g}=-\frac 2{g^3}\int \omega _D^0+B_\Sigma
\Delta \,^{-1},  \label{top}
\end{equation}
where

\begin{equation}
\omega _D^0=d^Dx{\cal L}_{{\rm inv}}\;+\;\left( \Phi ^{*}-{\rm %
dependent\;terms}\right) \;  \label{omo}
\end{equation}
is a nonintegrated field polynomial with form degree $D$ and zero ghost
number and $\Delta \,^{-1}$ is a trivial integrated cocycle with negative
ghost number. The appearance of possible antifields dependent terms in the
right-hand side of eq.$\left( \ref{omo}\right) $ accounts for the case in
which one has to deal with open gauge algebras, which close only up to
equations of motion. As we shall see, this will be the case of $N=2$ and $%
N=4 $ SYM.

Hence, the integrated consistency condition 
\begin{equation}
B_\Sigma \int \omega _D^0=0\;  \label{bcoh}
\end{equation}
can be translated at the nonintegrated level, giving rise to the following
set of descent equations \cite{book}

\begin{eqnarray}
B_\Sigma \,\omega _D^0+d\,\omega _{D-1}^1 &=&0\,\,,  \nonumber  \label{5} \\
B_\Sigma \,\omega _{D-1}^1+d\,\omega _{D-2}^2 &=&0\,\,,  \nonumber \\
&&...,  \nonumber \\
B_\Sigma \,\omega _1^{D-1}+d\,\omega _0^D &=&0\,\,,  \nonumber \\
B_\Sigma \,\omega _0^D &=&0\,\,,  \label{de}
\end{eqnarray}
with $\omega _{D-p\;}^p(p=0,...,D)$ being local field polynomials with form
degree ($D-p)$ and ghost number $p$.

In what follows we shall be interested in the class of models fulfilling the
two assumptions given below:

\begin{itemize}
\item  {\bf i)} \ The cohomology of $B_\Sigma $ is empty in all sectors with
form degree $1\leq p\leq D$ .

\item  {\bf ii)\ }The sector with form degree zero is nonvanishing, with a
unique{\it \ }nontrivial element $\omega _0^D$.
\end{itemize}

\subsection{Quantum aspects}

Concerning the quantum aspects, the first requirement is the absence of
anomalies in the quantum extension of the Slavnov-Taylor identity, {\it i.e.}
\begin{eqnarray}
\Gamma &=&\Sigma +O\left( \hbar \right) \,,  \nonumber  \label{9} \\
{\cal S}\left( \Gamma \right) &=&0\,,  \label{qst}
\end{eqnarray}
where $\Gamma $ is the 1PI\ generating functional.

As usual, the dependence of $\Gamma $ from the renormalization point $\mu $
is governed by the Callan-Symanzik equation, whose generic form reads

\begin{equation}
{\cal C\,}\Gamma =0\,\,\,,\,\;\;\,\,{\cal C}\equiv \mu \frac \partial
{\partial \mu }+\hbar \beta _g\frac \partial {\partial g}-\hbar \,\gamma
_{\Phi ^i}\,N_{\Phi ^i}\,,  \label{cs}
\end{equation}
where $\beta _g$ is the beta function, $\gamma _{\Phi ^i}$ stand for the
anomalous dimensions of the fields, and $N_{\Phi ^i}$ is the counting
operator

\begin{equation}
N_{\Phi ^i}=\int d^Dx\left( \Phi ^i\frac \delta {\delta \Phi ^i}-\Phi
^{i*}\frac \delta {\delta \Phi ^{i*}}\right) \;.  \label{count}
\end{equation}
Following the procedure outlined in ref.\cite{book} and making use of the
absence of anomalies in the Slavnov-Taylor identity $\left( \ref{qst}\right)
,$ the cocycles $\left\{ \omega _{D-p}^p;0\leq p\leq D\right\} $ can be
promoted to quantum insertions $\left[ \omega _{D-p}^p\cdot \Gamma \right]
\, $ fulfilling the quantum version of the descent equations $\left( \ref{de}%
\right) ,$ {\it i.e.}

\begin{eqnarray}
B_\Gamma \left[ \omega _{D-p}^p\cdot \Gamma \right] +d\left[ \omega
_{D-p-1}^{p+1}\cdot \Gamma \right] &=&0\,\,,  \nonumber \\
B_\Gamma \left[ \omega _0^D\cdot \Gamma \right] &=&0\,\,\,\,.  \label{qde}
\end{eqnarray}
As shown in \cite{book}, the insertions $\left[ \omega _{D-p}^p\cdot \Gamma
\right] $ possess the same anomalous dimension $\gamma _{\omega \text{ }}$%
and obey the following Callan-Symanzik equation

\begin{equation}
{\cal C}\left[ \omega _{D-p}^p\cdot \Gamma \right] +\hbar \gamma _{\omega 
\text{ }}\left[ \omega _{D-p}^p\cdot \Gamma \right] =\hbar \,B_\Gamma
\,\left[ \Xi _{D-p}^{p-1}\cdot \Gamma \right] \;,  \label{cs-omo}
\end{equation}
for some cohomologically trivial local polynomial $\Xi _{D-p}^{p-1}$.

The last important assumption which we shall require is that the anomalous
dimension $\gamma _{\omega \text{ }}$of the insertion $\left[ \omega
_0^D\cdot \Gamma \right] $ vanishes, {\it i.e. }$\gamma _{\omega \text{ }}=0$%
. Thus

\begin{equation}
{\cal C}\left[ \,\omega _0^D\cdot \Gamma \right] =\hbar \,B_\Gamma \,\left[
\Xi _0^{D-1}\cdot \Gamma \right] \;,  \label{ad}
\end{equation}
which, of course, implies that

\begin{equation}
{\cal C}\left[ \int \,\omega _D^0\cdot \Gamma \right] =\hbar \,B_\Gamma
\,\left[ \int \Xi _D^{-1}\cdot \Gamma \right] \;.  \label{cstop}
\end{equation}
In summary, we are dealing with a theory for which there exists a {\it one
to one} relationship between the solutions $\omega _D^0$ and $\omega _0^D$
corresponding to the top and to the bottom levels of the classical descent
equations $\left( \ref{de}\right) .$ In addition, besides the absence of
anomalies in the Slavnov-Taylor identity, the quantum insertion $\left[
\,\omega _0^D\cdot \Gamma \right] $ is required to have vanishing anomalous
dimension, as stated by eq.$\left( \ref{ad}\right) $. These features will
strongly constrain the beta function $\beta _g$. The main idea underlying
this construction is that of exploiting the {\it one to one }correspondence%
{\it \ }between $\partial \Sigma /\partial g$ and the cocycle $\omega _0^D,$
which is not renormalized. It turns out that the nonrenormalization
properties of $\omega _0^D$ affect directly all cocycles entering the
descent equations $\left( \ref{de}\right) ,$ including, in particular, $%
\partial \Sigma /\partial g$ and its anomalous dimension, which is nothing
but the beta function $\beta _g$.

\section{The algebraic criterion for the ultraviolet finiteness}

\subsection{The finiteness theorem}

The aim of this section is to cast the previous considerations into a
precise statement about the beta function. Let $\beta _g^{\left( n\right) }$
denote the contribution of order $\hbar ^n$ to the beta function $\beta _g$.
The theory is specified by a quantum vertex functional $\Gamma =\Sigma
+O\left( \hbar \right) ,\,$which fulfills all the above mentioned
assumptions, namely, the classical requirements{\it \ }{\bf i) }and{\bf \ ii)%
}, and the quantum properties encoded in eqs.$\left( \ref{qst}\right) $ and $%
\left( \ref{cstop}\right) $.

The following theorem holds

{\bf Theorem:\ }{\it If }$\,${\it the one-loop order contribution }$\beta
_g^{\left( 1\right) }${\it \ vanishes, i.e. }$\beta _g^{\left( 1\right)
}=0,\,${\it \ then }$\beta _g${\it \ vanishes to all orders of perturbation
theory.}

\vspace{5mm}

{\bf Proof:} In order to prove the theorem, let us first show that the
following identity is valid 
\begin{equation}
\frac{\partial \Gamma }{\partial g}=-\frac 2{g^3}\widetilde{a}\left[ \int
\,\omega _D^0\cdot \Gamma \right] +B_\Gamma \left[ \Delta ^{-1}\cdot \Gamma
\right] \;\,,  \label{rel}
\end{equation}
where $\left[ \Delta ^{-1}\cdot \Gamma \right] $ is an integrated insertion
with negative ghost number and $\widetilde{a}$ is a formal power series in $%
\hbar $ 
\begin{equation}
\widetilde{a}=\left( 1+\sum_{j=1}^\infty \hbar ^ja_j\right) \;.  \label{atil}
\end{equation}
Notice also that the coefficients $a_j$ are dimensionless since the theory
is considered to be massless.

Eq.$\left( \ref{rel}\right) $ is indeed easily established by induction in $%
\hbar \,.$ At the zeroth order it is obviously verified due to eq.$\left( 
\ref{top}\right) $. Let us suppose then that it holds at the order $\hbar ^n$%
, {\it i.e.}

\begin{equation}
\frac{\partial \Gamma }{\partial g}=-\frac 2{g^3}\left( 1+\sum_{j=1}^n\hbar
^ja_j\right) \left[ \int \,\omega _D^0\cdot \Gamma \right] +B_\Gamma \left[ 
\widehat{\Delta }^{-1}\cdot \Gamma \right] +\hbar ^{n+1}\Theta
_{n+1}+O\left( \hbar ^{n+2}\right) \,.  \label{reln}
\end{equation}
where, from the Quantum Action Principle \cite{book}, $\Theta _{n+1}$ is an
integrated local polynomial with ghost number zero which obeys the condition 
\begin{equation}
B_\Sigma \Theta _{n+1}=0\;,  \label{gn1}
\end{equation}
following from

\begin{equation}
B_\Gamma \frac{\partial \Gamma }{\partial g}=0,\,\,\,\,B_\Gamma B_\Gamma
=0\,.  \label{cond}
\end{equation}
Therefore, taking into account that the unique nontrivial cohomology class
of $B_\Sigma $ with the same quantum numbers of the action is $\int \omega
_D^0$, we get

\begin{equation}
\Theta _{n+1}=a_{n+1}\int \,\omega _D^0+B_\Sigma \widehat{\Theta }%
_{n+1\;}^{-1},  \label{prel}
\end{equation}
which establishes the validity of eq.$\left( \ref{rel}\right) $ at the order 
$\hbar ^{n+1},$ and hence to all orders by induction.

Now, coming back to the proof of the theorem, we act with the
Callan-Symanzik operator ${\cal C}$ on the eq.$\left( \ref{rel}\right) $.
Making use of eqs.$\left( \ref{cs}\right) $ and $\left( \ref{cstop}\right) ,$
and recalling the exact commutation relation

\begin{equation}
{\cal C}B_\Gamma -B_\Gamma {\cal C}=0\,,  \label{comm}
\end{equation}
we get the condition

\begin{equation}
\left[ {\cal C},\frac \partial {\partial g}\right] \Gamma =-\left( {\cal C}%
\left( \frac 2{g^3}\widetilde{a}\right) \right) \left[ \int \,\omega
_D^0\cdot \Gamma \right] +\hbar B_\Gamma \left[ \Omega ^{-1}\cdot \Gamma
\right] \,\,,  \label{ccs}
\end{equation}
for some irrelevant trivial insertion $\left[ \Omega ^{-1}\cdot \Gamma
\right] $ with negative ghost number. Working out the commutator in the
left-hand side and observing that the dimensionless coefficients $a_j$ do
not depend on $\mu ,$ we obtain 
\begin{equation}
\left( \left( \frac \partial {\partial g}\beta _g\right) \frac 2{g^3}%
\widetilde{a}+\beta _g\frac \partial {\partial g}\left( \frac 2{g^3}%
\widetilde{a}\right) \right) \left[ \int \omega _D^0\cdot \Gamma \right]
=B_\Gamma \left[ \widehat{\Omega }^{-1}\cdot \Gamma \right] \;,  \label{eqb}
\end{equation}
which, due to the fact that the insertion $\left[ \int \omega _D^0\cdot
\Gamma \right] $ cannot be written as a pure $B_\Gamma -$variation, finally
implies the condition

\begin{equation}
\left( \frac \partial {\partial g}\beta _g\right) \frac 2{g^3}\widetilde{a}%
+\beta _g\frac \partial {\partial g}\left( \frac 2{g^3}\widetilde{a}\right)
=0\;.  \label{pro}
\end{equation}
This equation expresses the content of the theorem, stating indeed that if
the one-loop contribution to the beta function vanishes, $\beta _g^{\left(
1\right) }=0,\,$ then $\beta _g=0$.

For a better understanding of the eq.$\left( \ref{pro}\right) $ let us
expand $\beta _g$ and $\widetilde{a}$ in powers of $\hbar $, yielding

{\bf order 1 : }

\begin{equation}
g\frac{\partial \beta _g^{(1)}}{\partial g}-3\beta _g^{(1)}=0\,\,\Rightarrow
\,\,\beta _g^{(1)}\sim g^3.  \label{o1}
\end{equation}

{\bf order 2 : }

\begin{equation}
\left( g\frac{\partial \beta _g^{(2)}}{\partial g}-3\beta _g^{(2)}\right)
+\beta _g^{(1)}\,g\frac{\partial a_1}{\partial g}=0.  \label{o2}
\end{equation}

{\bf order n :}

\begin{equation}
\left( g\frac{\partial \beta _g^{(n)}}{\partial g}-3\beta _g^{(n)}\right) +%
{\sum_{i=1}^{n-1}}\left( \left( g\frac{\partial
\beta _g^{(n-i)}}{\partial g}-3\beta _g^{(n-i)}\right) a_i+\beta
_g^{(n-i)}\,g\frac{\partial a_i}{\partial g}\right) =0.  \label{on}
\end{equation}
It becomes apparent thus that if $\beta _g^{(1)}=0$ in the above equations,
then $\beta _g^{(n)}=0$ for all $n$.

Before discussing the applications of this result, let us underline that the
present set up provides also a simple algebraic understanding of the case in
which $\beta _{g\text{ }}$receives contributions only up to one-loop order
as, for instance, in the $N=2$ SYM. This will be the aim of the next
subsection.

\subsection{Absence of higher order corrections}

It is known that the beta function $\beta _g$ depends on the renormalization
scheme, only the first order coefficient being universal \cite{weinberg}.
However, for some theories it happens that $\beta _g$ receives contributions
only up to one-loop order. This statement means really that there exist
renormalization schemes in which all the higher loop corrections vanish.
These schemes can be identified in an algebraic way by the following
proposition

{\bf Proposition: }For any renormalization scheme in which the following
identity holds

\begin{equation}
\frac{\partial \Gamma }{\partial g}=-\frac 2{g^3}\left[ \int \,\omega
_D^0\cdot \Gamma \right] +B_\Gamma \left[ \Delta ^{-1}\cdot \Gamma \right]
\,,  \label{lem}
\end{equation}
for some integrated insertion $\left[ \Delta ^{-1}\cdot \Gamma \right] $,
then $\beta _g$ has at most one-loop contributions.

\vspace{0.5cm}

{\bf Proof: }The equation $\left( \ref{lem}\right) $ is equivalent to $%
\left( \ref{rel}\right) $ with the requirement that now $a_j=0$ for any $j$.
Repeating therefore the same steps as before, the equation $\left( \ref{pro}%
\right) $ becomes

\begin{equation}
g\frac{\partial \beta _g}{\partial g}-3\beta _g=0\,\;,  \label{blem}
\end{equation}
which implies that {\it \ }$\beta _g$ has only one-loop contributions, {\it %
i.e.\ }$\beta _g\sim g^3.\;$The identity $\left( \ref{lem}\right) $ will
turn out to be very useful in the analysis of $N=2$ SYM.

\section{Applications}

In this section we shall present some applications of the finiteness
criterion discussed previously. Let us begin with the case of the
three-dimensional nonabelian Chern-Simons theory coupled to spinor matter.

\subsection{Chern-Simons coupled to matter}

The classical invariant action of the model is given by :

\begin{equation}
S_{{\rm inv}}=\int d^3x\,\left( \frac 1{2g^2}\varepsilon ^{\mu \nu \rho }\,%
{\rm Tr\,}\left( A_\mu \partial _\nu A_\rho +\frac 23A_\mu A_\nu A_\rho
\right) +i\overline{\Psi }\,\gamma ^\mu \,D_\mu \Psi \right) \;.
\label{csinv}
\end{equation}
The gauge field $A_\mu \,$ belongs to the adjoint representation of a
general compact Lie group $G$

\begin{equation}
A_\mu (x)=A_\mu ^a(x)\,\tau _a  \label{cslav}
\end{equation}
where the matrices $\tau _a\,$are the generators of the group, chosen to be
antihermitean

\begin{equation}
\left[ \tau _a,\tau _b\right] =f_{abc}\,\tau _{c\;},\;\;\;\;\;\;{\rm Tr}%
\,\tau _a\tau _b=\delta _{ab}\;.  \nonumber
\end{equation}
The matter fields belong to some finite representation of $G$, the
corresponding generators being denoted by $T_a$. Hence, for the covariant
derivative we have

\begin{equation}
D_\mu \Psi =(\partial _\mu +A_\mu ^aT_a)\Psi \;.  \label{cscd}
\end{equation}
Adopting the Landau condition, the gauge-fixing term reads

\begin{equation}
S_{{\rm gf}}=s\,{\rm Tr}\int d^3x\,\,\overline{c}\partial ^\mu A_\mu ={\rm Tr%
}\int d^3x\,\left( b\partial ^\mu A_\mu +\overline{c}\partial ^\mu D_\mu
c\right)  \label{csgf}
\end{equation}
where $c,\overline{c}$ and $b$ denote respectively the Faddeev-Popov ghost,
the antighost and the lagrangian multiplier, all of them in the same
representation as $A_\mu $. The BRST\ operator $s$ acts on the fields as
follows

\begin{eqnarray}
sA_\mu &=&-D_\mu c\,=-\left( \partial _\mu c+\left[ A_\mu ,c\right] \right) 
\nonumber  \label{csbrs} \\
sc &=&c^2  \nonumber \\
s\Psi &=&c^aT_a\,\Psi  \nonumber \\
s\overline{\Psi } &=&\overline{\Psi }\,T_a\,c^a  \nonumber \\
s\overline{c} &=&b\;\;  \nonumber \\
sb &=&0\;.  \label{csbrst}
\end{eqnarray}
Coupling now the nonlinear BRST\ transformations to the antifields $A_\mu
^{*}$, $c^{*},$ $\overline{\Psi }^{*}$, $\Psi ^{*}$

\begin{equation}
S_{{\rm ext}}=\int d^3x\,\left( {\rm Tr}\,\left( -A_\mu ^{*}D^\mu
c+c^{*}c^2\right) +\overline{\Psi }^{*}c^aT_a\,\Psi -\overline{\Psi }%
\,T_a\,c^a\,\Psi ^{*}\right) \;,  \label{csext}
\end{equation}
it turns out that the fully quantized classical action $\Sigma $

\begin{equation}
\Sigma =S_{{\rm inv}}+S_{{\rm gf}}+S_{{\rm ext}}  \label{csact}
\end{equation}
obeys the Slavnov-Taylor identity

\begin{equation}
{\cal S}(\Sigma )=\int d^3x\,\left( {\rm Tr}\left( \frac{\delta \Sigma }{%
\delta A_\mu ^{*}}\frac{\delta \Sigma }{\delta A^\mu }+\frac{\delta \Sigma }{%
\delta c^{*}}\frac{\delta \Sigma }{\delta c}+b\frac{\delta \Sigma }{\delta 
\overline{c}}\right) +\frac{\delta \Sigma }{\delta \overline{\Psi }^{*}}%
\frac{\delta \Sigma }{\delta \Psi }-\frac{\delta \Sigma }{\delta \Psi ^{*}}%
\frac{\delta \Sigma }{\delta \overline{\Psi }}\right) =0\;.  \label{csst}
\end{equation}
Accordingly, the nilpotent linearized operator $B_\Sigma $ is given by

\begin{eqnarray}
B_\Sigma &=&\int d^3x\,\left( {\rm Tr}\left( \frac{\delta \Sigma }{\delta
A_\mu ^{*}}\frac \delta {\delta A^\mu }+\frac{\delta \Sigma }{\delta A^\mu }%
\frac \delta {\delta A_\mu ^{*}}+\frac{\delta \Sigma }{\delta c^{*}}\frac
\delta {\delta c}+\frac{\delta \Sigma }{\delta c}\frac \delta {\delta
c^{*}}+b\frac \delta {\delta \overline{c}}\right) \right.  \nonumber
\label{csst} \\
&&\left. +\frac{\delta \Sigma }{\delta \overline{\Psi }^{*}}\frac \delta
{\delta \Psi }+\frac{\delta \Sigma }{\delta \Psi }\frac \delta {\delta 
\overline{\Psi }^{*}}-\frac{\delta \Sigma }{\delta \Psi ^{*}}\frac \delta
{\delta \overline{\Psi }}-\frac{\delta \Sigma }{\delta \overline{\Psi }}%
\frac \delta {\delta \Psi ^{*}}\right) \;.  \label{csstl}
\end{eqnarray}
For further use, the quantum numbers of all fields and antifields are
displayed in Table 1.,

\begin{eqnarray*}
&& 
\begin{tabular}{|l|l|l|l|l|l|l|l|l|l|l|}
\hline
& $A_\mu $ & $c$ & $\overline{c}$ & $b$ & $\overline{\Psi }$ & $\Psi $ & $%
A_\mu ^{*}$ & $c^{*}$ & $\Psi ^{*}$ & $\overline{\Psi }^{*}$ \\ \hline
Dim. & 1 & 0 & 1 & 1 & 1 & 1 & 2 & 3 & 2 & 2 \\ \hline
N.Ghost & 0 & 1 & -1 & 0 & 0 & 0 & -1 & -2 & -1 & -1 \\ \hline
Nature & C & A & A & C & A & A & A & C & C & C \\ \hline
\end{tabular}
\\
&&_{\,\,\,\,\,\,\,\,\,\,\,\,\,\,\,\,\,\,\text{Table 1. Dimension, ghost
number and nature of the fields. }}
\end{eqnarray*}
Having quantized the theory, let us turn to the characterization of the
cohomology of $B_\Sigma $ in the sector of the invariant counterterms

\begin{equation}
B_\Sigma \Delta ^0=0\;,  \label{cscc}
\end{equation}
where $\Delta ^0$ is an integrated local polynomial with dimension three and
zero ghost number. Setting

\begin{equation}
\Delta ^0=\int d^3x\,\omega ^0\;,  \label{csct}
\end{equation}
we obtain the following set of descent equations

\begin{eqnarray}
B_\Sigma \omega ^0 &=&\partial ^\mu \omega _\mu ^1\,,  \nonumber \\
B_\Sigma \omega _\mu ^1 &=&\partial ^\nu \omega _{\left[ \mu \nu \right]
}^2\,,  \nonumber \\
B_\Sigma \omega _{\left[ \mu \nu \right] }^2 &=&\partial ^\rho \omega
_{\left[ \mu \nu \rho \right] }^3\,,  \nonumber \\
B_\Sigma \omega _{\left[ \mu \nu \rho \right] }^3 &=&0\;.  \label{csde}
\end{eqnarray}
The unique nontrivial solution for $\omega _{\left[ \mu \nu \rho \right] }^3$
is given by

\begin{equation}
\omega _{\left[ \mu \nu \rho \right] }^3=\varsigma \,\varepsilon _{\mu \nu
\rho }\,\frac 13{\rm Tr\,}c^3  \label{csbl}
\end{equation}
where $\varsigma $ is a constant parameter. The higher cocycles $\omega
^0,\omega _\mu ^1$ and $\omega _{\left[ \mu \nu \right] }^2$ are easily
worked out and found to be 
\begin{eqnarray}
\omega _{\left[ \mu \nu \right] }^2 &=&-\varsigma \,\varepsilon _{\mu \nu
\rho }\,{\rm Tr\,}c\partial ^\rho c\,,  \nonumber \\
\omega _\mu ^1 &=&\varsigma \,\varepsilon _{\mu \nu \rho }\,{\rm Tr\,}A^\nu
\partial ^\rho c\,,  \nonumber \\
\omega ^0 &=&-\varsigma \,\varepsilon ^{\mu \nu \rho }\,{\rm Tr\,}\left(
A_\mu \partial _\nu A_\rho +\frac 23A_\mu A_\nu A_\rho \right) \;.
\label{cslev}
\end{eqnarray}
Concerning possible contributions coming from the spinor fields and the
antifields, it turns out by explicit inspection that they give rise only to
cohomologically trivial solutions, as can be straightforwardly checked with
the Dirac term appearing in the complete action $\Sigma $, namely

\begin{equation}
i\overline{\Psi }\gamma ^\mu D_\mu \Psi =B_\Sigma (\overline{\Psi }\Psi
^{*})\;.  \label{csdt}
\end{equation}
The solution given in eqs.$\left( \ref{csbl}\right) $ and $\left( \ref{cslev}%
\right) $ is thus the most general nontrivial solution of the descent
equations $\left( \ref{csde}\right) $. Of course, one has always the freedom
of adding trivial terms.

Acting now with $\partial /\partial g$ on the Slavnov-Taylor identity one
obtains

\begin{equation}
B_\Sigma \frac{\partial \Sigma }{\partial g}=0  \label{cscoh}
\end{equation}
with

\begin{equation}
\frac{\partial \Sigma }{\partial g}=-\frac 1{g^3}\int d^3x\,\varepsilon
^{\mu \nu \rho }\,{\rm Tr\,}\left( A_\mu \partial _\nu A_\rho +\frac 23A_\mu
A_\nu A_\rho \right)  \label{cssol}
\end{equation}
It becomes apparent therefore that $\partial \Sigma /\partial g$ coincides
with $\Delta ^0$ by taking $\varsigma =1/g^3.\;$In particular, $\partial
\Sigma /\partial g$ identifies the unique nontrivial class of the cohomology
of $B_\Sigma $ in the sector of counterterms. Moreover, there exists a {\it %
one to one} relationship between $\partial \Sigma /\partial g$ and the ghost
polynomial ${\rm Tr\,}c^3,$ implying that all classical assumptions of the
finiteness criterion are fulfilled. Concerning now the quantum aspects, we
point out that the Slavnov-Taylor identity can be established for the vertex
functional $\Gamma $, due to the well known absence of gauge anomaly in
three dimensions \cite{book}.

According then to the general set up, the last requirement to be satisfied
in order to apply the finiteness theorem is to prove that the gauge
invariant field polynomial ${\rm Tr\,}c^3$ can be promoted to a quantum
insertion $\left[ {\rm Tr\,}c^3\cdot \Gamma \right] $ with vanishing
anomalous dimension. This is ensured by the so called ghost equation Ward
identity \cite{ghe,book}

\begin{equation}
\int d^3x\,\left( \frac \delta {\delta c}+[\overline{c},\frac \delta {\delta
b}]\right) \Sigma =\Delta ^{{\rm cl}}\;,  \label{csge}
\end{equation}
where $\Delta ^{{\rm cl}}$ is a classical breaking

\begin{equation}
\Delta ^{{\rm cl}}=\int d^3x\left( [A_\mu ^{*},A^\mu ]-\left[ c^{*},c\right]
+\left( \overline{\Psi }^{*}T^a\,\Psi +\overline{\Psi }\,T^a\,\Psi
^{*}\right) \tau _a\right) .  \label{csdcl}
\end{equation}
As shown in detail in \cite{ghe,book} the Ward identity $\left( \ref{csge}%
\right) $ allows to control the dependence of the theory from the
Faddeev-Popov ghost, implying, in particular, the vanishing of the anomalous
dimension of $\left[ {\rm Tr\,}c^3\cdot \Gamma \right] $ to all orders of
perturbation theory.

Concerning the one-loop behavior of the beta function, it is worth reminding
here that the ultraviolet finiteness of Chern-Simons at one-loop order, with
or without matter, is a well known result, being checked in many ways by
several authors (see for instance \cite{csl}). Therefore, according to the
finiteness theorem, $\beta _g\;$vanishes to all orders of perturbations
theory. This example shows in a rather simple way that a great amount of
information on the beta function $\beta _g$ follows from the knowledge of
the anomalous dimension of the gauge invariant insertion $\left[ {\rm Tr\,}%
c^3\cdot \Gamma \right] .$

\subsection{$N=2$ Super Yang-Mills}

The nonrenormalization theorem of the beta function of $N=2$ SYM, stating
that $\beta _g$ receives only one-loop contributions, has long been known 
\cite{n2l,n2gri}. Recently, a purely algebraic proof of this result, based
on BRST\ Ward identities, has been given in \cite{n2sor}. It can be
considered as a highly nontrivial realization of the algebraic finiteness
criterion. In this subsection we shall review the main steps of the proof
within the present context.

In order to study the quantum properties of $N=2$ we shall make use of the
twisting procedure which allows to replace the spinor indices of
supersymmetry $(\alpha ,\dot{\alpha})$ with Lorentz vector indices. The
physical content of the theory is left unchanged, since the twist is a
linear change of variables, and the twisted version is perturbatively
indistinguishable from the original one. However, the use of the twisted
variables considerably simplifies the analysis of the finiteness properties,
allowing to identify a subset of supercharges which is actually relevant to
control the ultraviolet behavior.

Let us begin by sketching the twisting procedure for the $N=2$ SYM in the
Wess-Zumino (WZ) gauge \cite{n2outros,n2sor}. The global symmetry group of $%
N=2$ in four dimensional flat euclidean space-time is $SU(2)_L\times
SU(2)_R\times SU(2)_I\times U(1)_{{\cal R}}$, where $SU(2)_L\times SU(2)_R$
is the rotation group and $SU(2)_I$ and $U(1)_{{\cal R}}$ are the symmetry
groups corresponding to the internal $SU(2)$-transformations and to the $%
{\cal R}$-symmetry. The twisting procedure consists of replacing the
rotation group by $SU(2)_L\times SU(2)_R^{\prime }$, where $SU(2)_R^{\prime
} $ is the diagonal sum of $SU(2)_R$ and $SU(2)_I$, allowing to identify the
internal indices with the spinor indices. The fields of the $N=2$ vector
multiplet in the WZ gauge are given by $(A_\mu ,\psi _\alpha ^i,\overline{%
\psi }_{\dot{\alpha}}^i,\phi ,\overline{\phi })$, where $\psi _\alpha ^i,%
\overline{\psi }_{\dot{\alpha}}^i$ are Weyl spinors with $i=1,2$ being the
internal index of the fundamental representation of $SU(2)_I,$ and $\phi ,%
\overline{\phi }$ are complex scalars. All fields belong to the adjoint
representation of the gauge group. Under the twisted group, these fields
decompose as \cite{n2sor,n2outros} 
\begin{eqnarray}
A_\mu &\rightarrow &A_\mu ,\,\,\,\,\,\,\,\hspace{1cm}(\phi ,\overline{\phi }%
)\rightarrow (\phi ,\overline{\phi })  \nonumber \\
\psi _\alpha ^i &\rightarrow &(\eta ,\;\chi _{\mu \nu }),\,\,\,\,\,%
\hspace{1cm}\overline{\psi }_{\dot{\alpha}}^i\rightarrow \psi _\mu .
\label{n2tw}
\end{eqnarray}
Notice that $(\psi _\mu ,\chi _{\mu \nu },\eta )$ anticommute due to their
spinor nature, and $\chi _{\mu \nu }$ is a self-dual tensor field. The
action of $N=2$ SYM in terms of the twisted variables is found to be \cite
{n2sor,n2outros} 
\begin{eqnarray}
&&S^{{\rm N=2}}=\frac 1{g^2}{\rm Tr}\int d^4x\left( \frac 12F_{\mu \nu
}^{+}F^{+\mu \nu }+\frac 12\overline{\phi }\left\{ \psi ^\mu ,\psi _\mu
\right\} -\chi ^{\mu \nu }(D_\mu \psi _\nu -D_\nu \psi _\mu )^{+}\right. 
\nonumber \\
&&\left. \hspace{2mm}+\eta D_\mu \psi ^\mu -\frac 12\overline{\phi }D_\mu
D^\mu \phi -\frac 12\phi \left\{ \chi ^{\mu \nu },\chi _{\mu \nu }\right\}
-\frac 18\left[ \phi ,\eta \right] \eta -\frac 1{32}\left[ \phi ,\overline{%
\phi }\right] \left[ \phi ,\overline{\phi }\right] \right) \;,  \nonumber
\label{n2ac} \\
&&  \label{n2ac}
\end{eqnarray}
where $g$ is the {\it unique} coupling constant and 
\begin{eqnarray}
&&F_{\mu \nu }^{+}=F_{\mu \nu }+\frac 12\epsilon _{\mu \nu \rho \sigma
}F^{\rho \sigma }\;\;\;,\;\;\;\tilde{F_{\mu \nu }}^{+}=\frac 12\epsilon
_{\mu \nu \rho \sigma }F^{+\rho \sigma }=F_{\mu \nu }^{+}\;,  \nonumber
\label{n2fie} \\
&&(D_\mu \psi _\nu -D_\nu \psi _\mu )^{+}=(D_\mu \psi _\nu -D_\nu \psi _\mu
)+\frac 12\epsilon _{\mu \nu \rho \sigma }(D^\rho \psi ^\sigma -D^\sigma
\psi ^\rho )\;.  \nonumber  \label{n2self} \\
&&  \label{n2self}
\end{eqnarray}
Also, it is easily seen that assigning to $\left( A_\mu ,\psi _\mu ,\chi
_{\mu \nu },\eta ,\phi ,\overline{\phi }\right) $ the following ${\cal R}$%
-charges $(0,-1,1,-1,2,-2)$, the expression $\left( \ref{n2ac}\right) $ has
vanishing total ${\cal R}$-charge.

The action $S^{{\rm N=2}}$ is invariant under gauge transformations with
infinitesimal parameter $\zeta $ 
\begin{eqnarray}
&&\delta _\zeta ^gA_\mu =-D_\mu \zeta =-\left( \partial _\mu \zeta +\left[
A_\mu ,\zeta \right] \right) ,  \nonumber \\
&&\delta _\zeta ^g\gamma =\left[ \zeta ,\gamma \right] \;\;,\;{\rm {with}%
\;\;\gamma =\left( \psi _\mu ,\chi _{\mu \nu },\eta ,\phi ,\overline{\phi }%
\right) .}  \label{n2gau}
\end{eqnarray}
which lead to the usual BRST transformations, with $\delta _\zeta
^g\rightarrow s$ and $\zeta \rightarrow c$, where $c$ is the Faddeev-Popov
ghost transforming as $sc=c^2$.

Concerning the supersymmetry generators $(\delta _i^\alpha ,\overline{\delta 
}_{\dot{\alpha}}^i)$ of the $N=2$ superalgebra, it turns out that the
twisting procedure gives rise to the following twisted generators: a scalar $%
\delta $, a vector $\delta _\mu $ and a self-dual tensor $\delta _{\mu \nu }$%
, which of course leave the action invariant. It is worth emphasizing that $%
S^{{\rm N=2}}$ is uniquely fixed by the scalar $\delta $ and the vector $%
\delta _\mu $ twisted charges. Due to this property, the tensor generator $%
\delta _{\mu \nu }$ will not be taken into further account, although its
inclusion can be done straightforwardly.

In order to properly quantize the theory we collect all the generators $%
(s,\delta ,\delta _\mu )$ into an extended operator $Q$, which turns out to
be nilpotent on-shell and modulo the space-time translations 
\begin{eqnarray}
Q=s+\omega \delta +\varepsilon ^\mu \delta _\mu \ ,  \label{n2q}
\end{eqnarray}
\begin{eqnarray}
Q^2=0{\ }+\omega \varepsilon ^\mu \partial _\mu +{\rm eqs.{\ }of{\ }motion}\
,  \label{n2qn}
\end{eqnarray}
where $\omega $ and $\varepsilon ^\mu $ are global ghosts. The operator $Q$
acts on the fields as 
\begin{eqnarray}
&&QA_\mu =-D_\mu c+\omega \psi _\mu +\frac{\varepsilon ^\nu }2\chi _{\nu \mu
}+\frac{\varepsilon _\mu }8\eta ,  \nonumber \\
&&Q\psi _\mu =\left\{ c,\psi _\mu \right\} -\omega D_\mu \phi +\varepsilon
^\nu \left( F_{\nu \mu }-\frac 12F_{\nu \mu }^{+}\right) -\frac{\varepsilon
_\mu }{16}\left[ \phi ,\bar{\phi}\right] ,  \nonumber \\
&&Q\chi _{\sigma \tau }=\left\{ c,\chi _{\sigma \tau }\right\} +\omega
F_{\sigma \tau }^{+}+\frac{\varepsilon ^\mu }8\left( \epsilon _{\mu \sigma
\tau \nu }+g_{\mu \sigma }g_{\nu \tau }-g_{\mu \tau }g_{\nu \sigma }\right)
D^\nu \bar{\phi},  \nonumber \\
&&Q\eta =\left\{ c,\eta \right\} +\frac \omega 2\left[ \phi ,\bar{\phi}%
\right] +\frac{\varepsilon ^\mu }2D_\mu \bar{\phi},  \nonumber \\
&&Q\phi =\left[ c,\phi \right] -\varepsilon ^\mu \psi _\mu ,  \nonumber \\
&&Q\bar{\phi}=\left[ c,\bar{\phi}\right] +2\omega \eta ,  \nonumber \\
&&Qc=c^2-\omega ^2\phi -\omega \varepsilon ^\mu A_\mu +\frac{\varepsilon ^2}{%
16}\bar{\phi},  \label{n2qtr}
\end{eqnarray}
Following the Batalin-Vilkovisky procedure, for the complete gauge-fixed
action we obtain \cite{n2sor,n2outros} 
\begin{eqnarray}
\Sigma =S^{{\rm N=2}}+S_{{\rm gf}}+S_{{\rm ext}},  \label{n2totac}
\end{eqnarray}
where $S_{{\rm gf}}$ is the gauge-fixing term in the Landau gauge and $S_{%
{\rm ext}}$ contains the coupling of the non-linear transformations $Q\Phi _i
$ to antifields $\Phi _i^{*}=(A_\mu ^{*}$, $\psi _\mu ^{*}$, $\frac 12\chi
_{\mu \nu }^{*}$, $\eta ^{*}$, $\phi ^{*}$, $\overline{\phi }^{*}$, $c^{*})$%
. They are given by\footnote{%
The presence of terms quadratic in the antifields in $S_{{\rm ext}}$ is due
to the fact that the operator $Q$ is nilpotent up to the equations of motion.%
} 
\begin{eqnarray}
&&S_{{\rm gf}}=Q\int d^4x{\rm Tr}\left( \bar{c}\partial A\right) {\rm ,} 
\nonumber \\
S_{{\rm ext}} &=&{\rm Tr}\int d^4x\left( \Phi _i^{*}Q\Phi _i+\frac{g^2}{32}%
\left( 4\omega ^2\chi ^{*2}-8\omega \varepsilon _\mu \chi ^{*\mu \nu }\psi
_\nu ^{*}+\varepsilon ^2\psi ^{*2}-(\varepsilon \psi ^{*})^2\right) \right) ,
\nonumber  \label{n2acs} \\
&&  \label{n2acs}
\end{eqnarray}
with 
\begin{eqnarray}
Q\bar{c}=b,\;\;\;\;\;\;Qb=\omega \varepsilon ^\mu \partial _\mu \bar{c},
\label{n2qb}
\end{eqnarray}
where, as usual, $\bar{c},b$ denote the antighost and the Lagrange
multiplier.

The complete action $\Sigma $ satisfies thus the following Slavnov-Taylor
identity 
\begin{eqnarray}
{\cal S}(\Sigma )=\omega \varepsilon ^\mu \Delta _\mu ^{{\rm cl}},
\label{n2slav}
\end{eqnarray}
where 
\begin{eqnarray}
{\cal S}(\Sigma )={\rm Tr}\int d^4x\left( \frac{\delta \Sigma }{\delta \Phi
_i^{*}}\frac{\delta \Sigma }{\delta \Phi _i}+b\frac{\delta \Sigma }{\delta 
\bar{c}}+\omega \varepsilon ^\mu \partial _\mu \bar{c}\frac{\delta \Sigma }{%
\delta b}\right) {\rm .}  \label{n2st}
\end{eqnarray}
and $\Delta _\mu ^{{\rm cl}}$ is an integrated local polynomial 
\begin{eqnarray}
\Delta _\mu ^{{\rm cl}} &=&{\rm Tr}\int d^4x\left( \frac {}{}c^{*}\partial
_\mu c-\phi ^{*}\partial _\mu \phi -A^{*\nu }\partial _\mu A_\nu +\psi
^{*\nu }\partial _\mu \psi _\nu \right.  \nonumber \\
&&\left. \hspace{4mm}-\bar{\phi}^{*}\partial _\mu \bar{\phi}+\eta
^{*}\partial _\mu \eta +\frac 12\chi ^{*\nu \rho }\partial _\mu \chi _{\nu
\rho }\right) {\rm .}  \label{n2cb}
\end{eqnarray}
Notice that $\Delta _\mu ^{{\rm cl}}$, being linear in the quantum fields,
is a classical breaking and will not be affected by the quantum corrections.
From the Slavnov-Taylor identity it follows that the linearized operator $%
B_\Sigma $ defined as 
\begin{eqnarray}
B_\Sigma ={\rm Tr}\int d^4x\left( \frac{\delta \Sigma }{\delta \Phi _i^{*}}%
\frac \delta {\delta \Phi _i}+\frac{\delta \Sigma }{\delta \Phi _i}\frac
\delta {\delta \Phi _i^{*}}+b\frac \delta {\delta \bar{c}}+\omega
\varepsilon ^\mu \partial _\mu \bar{c}\frac \delta {\delta b}\right)
\label{n2lst}
\end{eqnarray}
turns out to be nilpotent modulo a total space-time derivative, namely 
\begin{eqnarray}
B_\Sigma B_\Sigma =\omega \varepsilon ^\mu \partial _\mu .  \label{n2snil}
\end{eqnarray}
The appearance of the space-time translation operator $\partial _\mu $ in
the right-hand of eq.$\left( \ref{n2snil}\right) $ is due to the
supersymmetric structure of the theory. Of course, the operator $B_\Sigma $
can be considered nilpotent when acting on the space of the integrated local
polynomials. Moreover, as we shall see in detail, the presence of the
space-time derivative $\partial _\mu $ will give rise to a set of
nonstandard descent equations which will turn out to constrain very strongly
the possible nontrivial invariant counterterms. We will also be able to
prove that these equations can be solved in a systematic way by using the
twisted $N=2$ supersymmetric algebra.

Proceeding as in the previous example, we act with the operator $\partial
/\partial g$ on both sides of the Slavnov-Taylor identity $\left( \ref
{n2slav}\right) .$ Observing then that the linear breaking term $\Delta _\mu
^{{\rm cl}}$ does not depend on the coupling constant $g$, we get the
condition

\begin{equation}
B_\Sigma \frac{\partial \Sigma }{\partial g}=0\;,  \label{n2cond}
\end{equation}
which shows that $\partial \Sigma /\partial g$ is invariant under the action
of $B_\Sigma $. It remains to prove that $\partial \Sigma /\partial g$ is
nontrivial. We are led then to solve the consistency condition for the
integrated invariant counterterms 
\begin{equation}
B_\Sigma \int d^4x\,\Omega ^0=0\;,  \label{w-z}
\end{equation}
where $\Omega ^0$ has the same quantum numbers of the classical action of $%
N=2$. Due to eq.$\left( \ref{n2snil}\right) $, the integrated consistency
condition $\left( \ref{w-z}\right) $ can be translated at the local level as 
\begin{equation}
B_\Sigma \Omega ^0=\partial ^\mu \Omega _{\mu \;}^1,  \label{1-l}
\end{equation}
where $\Omega _{\mu \;}^1$is a local polynomial with ghost number 1 and
dimension 3. Applying now the operator $B_\Sigma $ to both sides of $\left( 
\ref{1-l}\right) $ and making use of eq.$\left( \ref{n2snil}\right) $, one
obtains the condition 
\begin{equation}
\partial ^\mu \left( B_\Sigma \Omega _{\mu \;}^1-\omega \varepsilon _\mu
\Omega ^0\right) =0\;,  \label{l-2}
\end{equation}
which, due to the algebraic Poincar\'{e} Lemma \cite{book}, implies 
\begin{equation}
B_\Sigma \Omega _{\mu \;}^1=\omega \varepsilon _\mu \Omega ^0+\partial ^\nu
\Omega _{[\nu \mu ]}^2\;,  \label{l-3}
\end{equation}
for some local polynomial $\Omega _{[\nu \mu ]}^2$ antisymmetric in the
Lorentz indices $\mu ,\nu $ and with ghost number 2. The procedure can be
easily iterated, yielding the following set of descent equations 
\begin{eqnarray}
B_\Sigma \Omega ^0 &=&\partial ^\mu \Omega _\mu ^1\;,  \nonumber \\
\,\,\,\,\,\,\,\,\;\;\;\;\;B_\Sigma \Omega _\mu ^1 &=&\partial ^\nu \Omega
_{[\nu \mu ]}^2+\omega \varepsilon _\mu \Omega ^0\;,  \nonumber \\
B_\Sigma \Omega _{[\mu \nu ]}^2 &=&\partial ^\rho \Omega _{[\rho \mu \nu
]}^3+\omega \varepsilon _\mu \Omega _\nu ^1-\omega \varepsilon _\nu \Omega
_\mu ^1\;,  \nonumber \\
B_\Sigma \Omega _{[\mu \nu \rho ]}^3 &=&\partial ^\sigma \Omega _{[\sigma
\mu \nu \rho ]}^4+\omega \varepsilon _\mu \Omega _{[\nu \rho ]}^2+\omega
\varepsilon _\rho \Omega _{[\mu \nu ]}^2+\omega \varepsilon _\nu \Omega
_{[\rho \mu ]}^2,  \nonumber \\
B_\Sigma \Omega _{[\mu \nu \rho \sigma ]}^4 &=&\omega \varepsilon _\mu
\Omega _{[\nu \rho \sigma ]}^3-\omega \varepsilon _\sigma \Omega _{[\mu \nu
\rho ]}^3+\omega \varepsilon _\rho \Omega _{[\sigma \mu \nu ]}^3-\omega
\varepsilon _\nu \Omega _{[\rho \sigma \mu ]}^3\,.  \nonumber  \label{DesEqB}
\\
&&  \label{DesEqB}
\end{eqnarray}
We observe that these equations are of a nonstandard type, as the cocycles
with lower ghost number appear in the equations of those with higher ghost
number, turning the system $\left( \ref{DesEqB}\right) $ nontrivial. We
remark that the last equation for $\Omega _{[\mu \nu \rho \sigma ]}^4$ is
not homogeneous, a property which strongly constrains the possible
solutions. Actually, it is possible to solve the system $\left( \ref{DesEqB}%
\right) $ in a rather direct way by making use of the $N=2$ structure. To
this end we introduce the operator 
\begin{equation}
{\cal W}_\mu =\frac 1\omega \left[ \frac \partial {\partial \varepsilon ^\mu
},B_\Sigma \right] \;,  \label{ClimbDesEq}
\end{equation}
which obeys the relations 
\begin{eqnarray}
\left\{ {\cal W}_\mu ,B_\Sigma \right\} &=&\partial _\mu \,,  \nonumber
\label{EBAlgebra} \\
\left\{ {\cal W}_\mu ,{\cal W}_\nu \right\} &=&0\,\;{\rm .}
\label{EBAlgebra}
\end{eqnarray}
This algebra is typical of topological quantum field theories \cite
{Delduc,ms}. In particular, as shown in \cite{Guadagnini}, the decomposition 
$\left( \ref{EBAlgebra}\right) $ allows to make use of ${\cal W}_\mu $ as a
climbing-up operator for the descent equations $\left( \ref{DesEqB}\right) .$
It turns out in fact that the nontrivial solution of the system is 
\begin{eqnarray}
\Omega ^0 &=&\frac 1{4!}{\cal W}^\mu {\cal W}^\nu {\cal W}^\rho {\cal W}%
^\sigma \Omega _{[\sigma \rho \nu \mu ]}^4\;,  \nonumber  \label{sol} \\
\Omega _\mu ^1 &=&\frac 1{3!}{\cal W}^\nu {\cal W}^\rho {\cal W}^\sigma
\Omega _{[\sigma \rho \nu \mu ]}^4\;,  \nonumber \\
\Omega _{[\mu \nu ]}^2 &=&\frac 1{2!}{\cal W}^\rho {\cal W}^\sigma \Omega
_{[\sigma \rho \mu \nu ]}^4\;,  \nonumber \\
\Omega _{[\mu \nu \rho ]}^3 &=&{\cal W}^\sigma \Omega _{[\sigma \mu \nu \rho
]}^4\;,  \label{sol}
\end{eqnarray}
with $\Omega _{[\mu \nu \rho \sigma ]}^4$ given by

\begin{equation}
\Omega _{[\mu \nu \rho \sigma ]}^4=\omega ^4\varepsilon _{\mu \nu \rho
\sigma }{\rm Tr}\phi ^2.  \label{nc}
\end{equation}
From eqs.$\left( \ref{sol}\right) $ the usefulness of the operator ${\cal W}%
_\mu $ becomes now apparent. Recalling thus that the cocycle $\Omega ^0$ has
the same quantum numbers of the $N=2$ Lagrangian, the following relation
holds 
\begin{equation}
\frac{\partial \Sigma }{\partial g}=\frac{2\omega ^4}{3g^3}\varepsilon ^{\mu
\nu \rho \sigma }{\cal W}_\mu {\cal W}_\nu {\cal W}_\rho {\cal W}_\sigma
\int d^4x{\rm Tr}\frac{\phi ^2}2\;+\;B_\Sigma \Xi ^{-1}\;,  \label{n2rel}
\end{equation}
for some irrelevant trivial $\Xi ^{-1}.$ This equation shows that there is a 
{\it one to one }relationship between the solution of the lowest level of
the descent equations $\left( \ref{DesEqB}\right) $ and the action of $N=2$,
so that the classical assumptions {\bf i) }and {\bf ii)} of Sect.2 are
satisfied. Equation $\left( \ref{n2rel}\right) $ implies that the
ultraviolet behavior of $N=2$ can be traced back to gauge invariant
polynomial ${\rm Tr}\phi ^2,$ which plays the r\^{o}le of a kind of
perturbative prepotential.

Concerning the quantum aspects, the Slavnov-Taylor identity $\left( \ref
{n2slav}\right) $ can be extended to the quantum level without anomalies 
\cite{nm}. Also, the construction given in \cite{book} can be generalized to
the set of descent equations $\left( \ref{DesEqB}\right) $, with the result
that the cocycles $\Omega ^0,$ $\Omega _\mu ^1,$ $\Omega _{[\mu \nu ]}^2,$ $%
\Omega _{[\mu \nu \rho ]}^3$, $\Omega _{[\mu \nu \rho \sigma ]}^4$ can be
promoted to quantum insertions with the same anomalous dimension. Finally,
the last requirement in order to apply the finiteness criterion is to
establish the vanishing of the anomalous dimension of the insertion $\left[ 
{\rm Tr}\phi ^2\cdot \Gamma \right] $. This important property has been
indeed proven in \cite{n2sor}. Without entering into further details, we
limit here to remark that the proof of the vanishing of the anomalous
dimension of $\left[ {\rm Tr}\phi ^2\cdot \Gamma \right] $ stems from a Ward
identity relating ${\rm Tr}\phi ^2$ to the gauge invariant polynomial ${\rm %
Tr}(-3\omega ^2c\phi +c^3)/\omega ^4$, whose anomalous dimension vanishes
due to the ghost equation \cite{n2sor}. In turn, this implies that $\left[ 
{\rm Tr}\phi ^2\cdot \Gamma \right] $ has vanishing anomalous dimension as
well. Moreover, in the present case, it has been possible to prove that the
classical equation $\left( \ref{n2rel}\right) $ can be extended as it stands
at the quantum level \cite{n2sor}, yielding the remarkable equation 
\begin{eqnarray}
\frac{\partial \Gamma }{\partial g}=\frac{2\omega ^4}{3g^3}\int d^4x\left[
\left( {\cal W}^4{\rm Tr}\frac{\phi ^2}2\right) \cdot \Gamma \right]
+B_\Gamma \left[ \Xi ^{-1}\cdot \Gamma \right] \;,  \label{rem}
\end{eqnarray}
with ${\cal W}^4$=$\varepsilon ^{\mu \nu \rho \sigma }{\cal W}_\mu {\cal W}%
_\nu {\cal W}_\rho {\cal W}_\sigma .$

We observe that this equation has the form of $\left( \ref{lem}\right) $,
implying, in particular, the absence of the coefficients $\widetilde{a}\;$of
eq.$\left( \ref{atil}\right) .$ Therefore, the proposition of subsect.3.2
applies with the result that the beta function of $N=2$ SYM is indeed of
one-loop order only, {\it i.e. }$\beta _g\sim g^3.$

\subsection{$N=4$ Super Yang-Mills}

The case of the $N=4$ SYM\ can be treated in a way similar to $N=2$. Let us
begin by describing how the twisting procedure can be applied. The global
symmetry group of $N=4$ SYM theory in euclidean space-time is $SU(2)_L\times
SU(2)_R\times SU(4)$, where $SU(2)_L\times SU(2)_R$ is the rotation group
and $SU(4)$ the internal symmetry group of $N=4$. Hence the twist operation
can be performed in more than one way, depending on how the internal
symmetry group is combined with the rotation group \cite{Yamron}. We shall
follow the procedure of Vafa and Witten \cite{Vafa}, in which the $SU(4)$ is
splitted as $SU(2)_F\times SU(2)_I$, so that the twisted global symmetry
group turns out to be $SU(2)_L^{\prime }\times SU(2)_R\times SU(2)_F,$ where 
$SU(2)_L^{\prime }={\rm diag}\left( SU(2)_L\oplus SU(2)_I\right) $ and $%
SU(2)_F$ is a residual internal symmetry group. The fields of the $N=4$
multiplet are given by $(A_\mu $,$\lambda _u^\alpha $,$\overline{\lambda }_{%
\dot{\alpha}}^u,\Phi _{uv})$, where $(u,v=1,..,4)$ are indices of the
fundamental representation of $SU(4)$, and the six real scalar fields of the
model are collected into the antisymmetric and self-conjugate tensor $\Phi
_{uv}$. Under the twisted group, these fields decompose as 
\begin{eqnarray}
A_\mu &\rightarrow &A_\mu \;,  \nonumber  \label{1-t} \\
\overline{\lambda }_{\dot{\alpha}}^u\; &\rightarrow &\psi _\mu ^i\;, 
\nonumber \\
\lambda _u^\alpha &\rightarrow &\eta ^i,\;\chi _{\mu \nu }^i\;,  \nonumber \\
\Phi _{uv} &\rightarrow &B_{\mu \nu },\;\phi ^{ij}\;,  \label{1-t}
\end{eqnarray}
where $(i,j=1,2)\;$are indices of the residual isospin group $SU(2)_F$, $%
\phi ^{ij}$ is a symmetric tensor, and $\chi _{\mu \nu }^i$,$\;B_{\mu \nu }$
are self-dual with respect to the Lorentz indices. Since in our analysis
manifest isospin invariance is not needed, we further explicit the $SU(2)_F$
doublets as $\psi _\mu ^i=(\psi _\mu ,\chi _\mu )$, $\eta ^i=(\eta ,\xi
),\;\chi _{\mu \nu }^i=(\chi _{\mu \nu },\psi _{\mu \nu })\;$and the triplet
as $\phi ^{ij}=(\phi ,\overline{\phi },\tau )$. The action of $N=4$ in terms
of the twisted fields is given by\footnote{%
The group generators are chosen here to be hermitian.} 
\begin{eqnarray}
S^{{\rm N=4}} &=&\frac 1{g^2}{\rm Tr}\int d^4x\left( \frac {}{}D_\mu \phi
D^\mu \overline{\phi }\,+\,i\psi _\mu D_\nu \chi ^{\mu \nu }\,+\,i\chi _\mu
D_\nu \psi ^{\mu \nu }-\,\chi _\mu D^\mu \xi \,\right.  \nonumber \\
&&+\,\psi _\mu D^\mu \eta {\rm \,}-\,i\overline{\phi }\left\{ \psi _{\mu \nu
},\psi ^{\mu \nu }\right\} \,+\,i\phi \left\{ \chi _{\mu \nu },\chi ^{\mu
\nu }\right\} \,+\,i\tau \left\{ \psi _{\mu \nu },\chi ^{\mu \nu }\right\} 
\nonumber \\
&&-\left\{ \psi _{\mu \nu },\chi ^{\mu \rho }\right\} B_\rho ^{\,\,\,\,\,\nu
}-i\chi _{\mu \nu }\left[ \xi ,B^{\mu \nu }\right] -i\psi _{\mu \nu }\left[
\eta ,B^{\mu \nu }\right] +4i\overline{\phi }\left\{ \xi ,\xi \right\} 
\nonumber \\
&&-\,\,4i\,\phi \,\left\{ \eta ,\eta \right\} \,\,+\,\,4i\,\tau \,\left\{
\xi ,\eta \right\} \,+\,\psi _\mu \,\left[ \chi _\nu ,B^{\mu \nu }\right]
\,\,+\,\,i\,\phi \,\left\{ \chi _\mu \,,\chi ^\mu \right\}  \nonumber \\
&&-i\overline{\phi }\,\left\{ \psi _\mu ,\psi ^\mu \right\} -\,i\psi _\mu
\,\left[ \chi ^\mu ,\tau \right] \,-4\,\left[ \phi ,\overline{\phi }\right]
\,\left[ \phi ,\overline{\phi }\right] \,+4\left[ \phi ,\tau \right] \left[ 
\overline{\phi },\tau \right]  \nonumber \\
&&+[\phi \,,B_{\mu \nu }]\,\,[\overline{\phi }\,,B^{\mu \nu }]\,-\,H^\mu
\,\left( \,H_\mu \,-\,D_\mu \tau \,+\,i\,D^\nu B_{\mu \nu }\,\right) 
\nonumber  \label{invaction} \\
&&\left. +H^{\mu \nu }\left( -H_{\mu \nu }+\frac i4F_{\mu \nu }^{+}-\frac
12\left[ B_{\mu \rho },B_{\,\,\,\nu }^\rho \right] -i\left[ B_{\mu \nu
},\tau \right] \right) \right) \,,  \label{invaction}
\end{eqnarray}
where $g$ is the unique coupling constant and $H_{\mu \nu }$, $H_\mu $ are
auxiliary fields, with $H_{\mu \nu }$ self-dual.

Concerning the generators $(\delta _u^\alpha ,\overline{\delta }_{\dot{\alpha%
}}^u)$ of the $N=4$ superalgebra, it turns out that the twisting procedure
gives rise to the following twisted charges \cite{Lozano}: two scalars, $%
\delta ^{+}$ and $\delta ^{-},$ two vectors, $\delta _\mu ^{+}$ and $\delta
_\mu ^{-}$, and two self-dual tensors $\delta _{\mu \nu }^{+}$ and $\delta
_{\mu \nu }^{-}.$ Of course, all twisted generators leave the action $\left( 
\ref{invaction}\right) $ invariant. It is worth emphasizing that, as proven
in \cite{n4sor}, the action $S^{{\rm N=4}}$ is uniquely fixed by the two
vector generators $\delta _\mu ^{+},\;\delta _\mu ^{-}$ and by the scalar
charge $\delta ^{+}.$ In other words, the requirement of invariance under $%
\delta _\mu ^{+},\;\delta _\mu ^{-}$ and $\delta ^{+{\rm \ }}$fixes all the
relative numerical coefficients of the various terms of the action $\left( 
\ref{invaction}\right) $. Thus, as done in the case of $N=2$, the tensorial
transformations $\delta _{\mu \nu }^{+}$, $\delta _{\mu \nu }^{-}$ will not
be taken into account. The action of the twisted $\delta ^{+}$ generator on
the fields reads: 
\begin{eqnarray}
&&
\begin{tabular}{ll}
$\delta ^{+}A_\mu =\psi _\mu \,,$ & $\delta ^{+}\tau =\xi $ \\ 
$\delta ^{+}\psi _\mu =D_\mu \phi \,,\,$ & $\delta ^{+}\chi _\mu =H_\mu \,,$
\\ 
$\delta ^{+}\phi =0\,\,$ & $\delta ^{+}\xi =i\left[ \tau ,\phi \right] $ \\ 
$\,\delta ^{+}\overline{\phi }=-\eta $ & $\delta ^{+}B_{\mu \nu }=\psi _{\mu
\nu }\,,$ \\ 
$\delta ^{+}\eta =i\left[ \phi ,\stackrel{\_}{\phi }\right] $ & $\delta
^{+}\psi _{\mu \nu }=i[B_{\mu \nu },\phi ]\,$ \\ 
$\delta ^{+}\chi _{\mu \nu }=H_{\mu \nu }$ & $\delta ^{+}H_\mu =i\left[ \chi
_\mu ,\phi \right] $ \\ 
$\delta ^{+}H_{\mu \nu }=i\left[ \chi _{\mu \nu },\phi \right] .$ & 
\end{tabular}
\nonumber \\
&&  \label{ScTransf+}
\end{eqnarray}
In the first column of eq.$\left( \ref{ScTransf+}\right) $ we recognize the
scalar transformations of the twisted $N=2$ subalgebra in presence of the
auxiliary field $H_{\mu \nu }.$ For $\delta ^{-}$ one gets 
\begin{eqnarray}
\! &&
\begin{tabular}{ll}
$\delta ^{-}A_\mu =\chi _\mu \,,$ & $\delta ^{-}\tau =-\eta \,,$ \\ 
$\delta ^{-}\chi _\mu =-D_\mu \overline{\phi }\,,$ & $\delta ^{-}\psi _\mu
=-H_\mu +D_\mu \tau \,,$ \\ 
$\,\delta ^{-}\stackrel{\_}{\phi }=0\,,$ & $\delta ^{-}\eta =i\left[ \tau ,%
\stackrel{\_}{\phi }\right] ,$ \\ 
$\delta ^{-}\phi =-\xi \,\,,$ & $\delta ^{-}\chi _{\mu \nu }=i\left[ B_{\mu
\nu },\stackrel{\_}{\phi }\right] ,$ \\ 
$\delta ^{-}\xi =i\left[ \phi ,\overline{\phi }\right] \,,$ & $\delta
^{-}B_{\mu \nu }=-\chi _{\mu \nu }\,,$ \\ 
$\delta ^{-}\psi _{\mu \nu }=H_{\mu \nu }+i\left[ B_{\mu \nu },\tau \right] ,
$ & 
\end{tabular}
\nonumber \\
&&
\begin{tabular}{l}
$\delta ^{-}H_{\mu \nu }=-i\left[ \psi _{\mu \nu },\stackrel{\_}{\phi }%
\right] +i\left[ \chi _{\mu \nu },\tau \right] +i\left[ B_{\mu \nu },\eta
\right] \;,$ \\ 
$\delta ^{-}H_\mu =-D_\mu \eta +i\left[ \psi _\mu ,\stackrel{\_}{\phi }%
\right] +i\left[ \chi _\mu ,\tau \right] \,\;.$%
\end{tabular}
\nonumber \\
&&  \label{ScTransf-}
\end{eqnarray}
Analogously, for the vector transformations $\delta _\mu ^{+}$ and $\delta
_\mu ^{-}$ one obtains 
\begin{eqnarray}
&&
\begin{tabular}{ll}
$\delta _\mu ^{+}A_\nu =-4i\chi _{\mu \nu }-4g_{\mu \nu }\eta \,,$ & $\delta
_\mu ^{+}\tau =\chi _\mu ,$ \\ 
$\delta _\mu ^{+}\phi =\psi _\mu \,,$ & $\delta _\mu ^{+}\stackrel{\_}{\phi }%
=0,$ \\ 
$\delta _\mu ^{+}\xi =D_\mu \tau -H_\mu \,,$ & $\delta _\mu ^{+}\eta =-D_\mu 
\overline{\phi },$ \\ 
$\delta _\mu ^{+}B_{\nu \rho }=-i\theta _{\mu \nu \rho \lambda }\chi
^\lambda \,,$ & $\delta _\mu ^{+}\psi _{\nu \rho }=D_\mu B_{\nu \rho
}+i\theta _{\mu \nu \rho \lambda }H^\lambda ,$ \\ 
$\delta _\mu ^{+}\chi _{\nu \rho }=i\theta _{\mu \nu \rho \lambda }D^\lambda 
\overline{\phi }\,,$ & $\delta _\mu ^{+}\chi _\nu =-4\left[ B_{\mu \nu },%
\overline{\phi }\right] +4ig_{\mu \nu }\left[ \tau ,\overline{\phi }\right] ,
$%
\end{tabular}
\nonumber \\
&&
\begin{tabular}{l}
$\delta _\mu ^{+}\psi _\nu =4iH_{\mu \nu }+F_{\mu \nu }-4ig_{\mu \nu }[%
\overline{\phi },\phi ]\,,$ \\ 
$\delta _\mu ^{+}H_{\nu \rho }=D_\mu \chi _{\nu \rho }+\theta _{\mu \nu \rho
\lambda }\left[ \psi ^\lambda ,\overline{\phi }\right] +i\theta _{\mu \nu
\rho \lambda }D^\lambda \eta \,,$ \\ 
$\delta _\mu ^{+}H_\nu =D_\mu \chi _\nu +4\left[ \eta ,B_{\mu \nu }\right]
+4\left[ \psi _{\mu \nu },\overline{\phi }\right] -4ig_{\mu \nu }\left[ \eta
,\tau \right] -4ig_{\mu \nu }\left[ \xi ,\overline{\phi }\right] \,,$%
\end{tabular}
\nonumber \\
&&  \label{VeTransf+}
\end{eqnarray}
and 
\begin{eqnarray}
&&
\begin{tabular}{ll}
$\delta _\mu ^{-}A_\nu =-4i\psi _{\mu \nu }+4g_{\mu \nu }\xi \,,$ & $\delta
_\mu ^{-}\tau =\psi _\mu ,$ \\ 
$\delta _\mu ^{-}\phi =0\,,$ & $\delta _\mu ^{-}\stackrel{\_}{\phi }=-\chi
_\mu ,$ \\ 
$\delta _\mu ^{-}\xi =-D_\mu \phi \,,$ & $\delta _\mu ^{-}\eta =-H_\mu ,$ \\ 
$\delta _\mu ^{-}B_{\nu \rho }=+i\theta _{\mu \nu \rho \lambda }\psi
^\lambda \,,$ & $\delta _\mu ^{-}\psi _\nu =-4\left[ B_{\mu \nu },\phi
\right] -4ig_{\mu \nu }\left[ \tau ,\phi \right] ,$ \\ 
$\delta _\mu ^{-}\psi _{\nu \rho }=-i\theta _{\mu \nu \rho \lambda
}D^\lambda \phi \,,$ & 
\end{tabular}
\nonumber \\
&&
\begin{tabular}{l}
$\delta _\mu ^{-}\chi _{\nu \rho }=-D_\mu B_{\nu \rho }-i\theta _{\mu \nu
\rho \lambda }H^\lambda +i\theta _{\mu \nu \rho \lambda }D^\lambda \tau \,,$
\\ 
$\delta _\mu ^{-}\chi _\nu =4iH_{\mu \nu }+F_{\mu \nu }+4ig_{\mu \nu }\left[ 
\overline{\phi },\phi \right] -4\left[ B_{\mu \nu },\tau \right] \,,$ \\ 
$\delta _\mu ^{-}H_{\nu \rho }=D_\mu \psi _{\nu \rho }+\theta _{\mu \nu \rho
\lambda }\left( \left[ \psi ^\lambda ,\tau \right] -\left[ \chi ^\lambda
,\phi \right] -iD^\lambda \xi \right) +i\left[ \psi _\mu ,B_{\nu \rho
}\right] \,,$ \\ 
$\delta _\mu ^{-}H_\nu =-D_\mu \psi _\nu +D_\nu \psi _\mu +4\left[ \psi
_{\mu \nu },\tau \right] -4\left[ \xi ,B_{\mu \nu }\right] +4\left[ \chi
_{\mu \nu },\phi \right] +4ig_{\mu \nu }\left[ \eta ,\phi \right] \,.$%
\end{tabular}
\nonumber \\
&&  \label{VeTransf-}
\end{eqnarray}
where $\theta _{\mu \nu \rho \sigma }$ denotes the combination 
\begin{equation}
\theta _{\mu \nu \rho \sigma }=\varepsilon _{\mu \nu \rho \sigma }+g_{\mu
\nu }g_{\rho \sigma }-g_{\mu \rho }g_{\nu \sigma }\;=4\Pi _{\mu \sigma \nu
\rho }^{+}\;,  \label{self}
\end{equation}
where $\Pi _{\mu \sigma \nu \rho }^{+}$ is the projector on self-dual
two-forms. Let us also give here the algebraic relations among the twisted
generators, {\it i.e. } 
\begin{eqnarray}
&&
\begin{tabular}{ll}
$\left\{ \delta ^{+},\delta ^{+}\right\} =\delta _{-2\phi }^g$ & $\left\{
\delta _\mu ^{+},\delta ^{+}\right\} =\partial _\mu +\delta _{A_\mu }^g$ \\ 
$\left\{ \delta ^{-},\delta ^{-}\right\} =\delta _{2\overline{\phi }}^g$ & $%
\left\{ \delta _\mu ^{-},\delta ^{-}\right\} =\partial _\mu +\delta _{A_\mu
}^g$ \\ 
$\left\{ \delta ^{+},\delta ^{-}\right\} =\delta _{-\tau }^g$ & $\left\{
\delta _\mu ^{+},\delta ^{-}\right\} =0$ \\ 
$\left\{ \delta _\mu ^{-},\delta ^{+}\right\} =0$ & $\left\{ \delta _\mu
^{+},\delta _\nu ^{+}\right\} =\delta _{-8g_{\mu \nu }\overline{\phi }}^g$
\\ 
$\left\{ \delta _\mu ^{-},\delta _\nu ^{-}\right\} =\delta _{8g_{\mu \nu
}\phi }^g$ & $\left\{ \delta _\mu ^{+},\delta _\nu ^{-}\right\} =\delta
_{-4iB_{\mu \nu }-4g_{\mu \nu }\tau }^g+{\rm eqs.\,\,of\,\,motion,}$%
\end{tabular}
\nonumber \\
&&  \label{DeltaAlgebra}
\end{eqnarray}
where $\delta _\gamma ^g$ denotes a gauge transformation with parameter $%
\gamma $.

In order to quantize the theory, we proceed as before and introduce a
generalized BRST\ operator $Q$ which collects all the symmetry generators 
\begin{equation}
Q=s+\omega ^{+}\delta ^{+}+\omega ^{-}\delta ^{-}+\varepsilon ^{+\mu }\delta
_\mu ^{+}+\varepsilon ^{-\mu }\delta _\mu ^{-}\;\,,  \label{QOperator}
\end{equation}
where $s$ is the ordinary BRST operator for the gauge transformations, and $%
\omega ^{+},\,\omega ^{-},\,\varepsilon ^{+\mu }$,$\,\varepsilon ^{-\mu }$
are global ghosts \cite{n4sor}. Defining the action of $Q$ on the
Faddeev-Popov ghost $c$ as

\begin{eqnarray}
Qc &=&ic^2+(\omega ^{+^2}-4\varepsilon ^{-^2})\phi +(4\varepsilon
^{+^2}-\omega ^{-^2})\overline{\phi }+(\omega ^{+}\omega ^{-}+4\varepsilon
^{+\mu }\varepsilon _\mu ^{-})\tau   \nonumber \\
&&+4i\varepsilon ^{+\mu }\varepsilon ^{-\nu }B_{\mu \nu }-(\omega
^{+}\varepsilon ^{+\mu }+\omega ^{-}\varepsilon ^{-\mu })A_\mu \;,
\label{q-c}
\end{eqnarray}
it follows that the operator $Q$ turns out to be nilpotent on shell and
modulo a total derivative 
\begin{equation}
Q^2=0{\ }+\left( \omega ^{+}\varepsilon ^{+\mu }\text{ }+\omega
^{-}\varepsilon ^{-\mu }\right) \partial _\mu +{\rm eqs.{\ }of{\ }motion}\ .
\label{qqn4}
\end{equation}
Introducing then a set of antifields $\Phi _i^{*}$ coupled to the nonlinear
transformations of the fields $Q\Phi _i,$ for the external action we obtain 
\begin{eqnarray}
S_{{\rm ex}\text{{\rm t}}} &=&{\rm Tr}\int d^4x\left( \;\Phi _i^{*}Q\Phi
_i\;+4g^2\varepsilon ^{+\mu }\varepsilon ^{-\nu }\left( \varepsilon _{\mu
\nu \rho \lambda }A^{*\rho }H^{*\lambda }\right. \right.   \nonumber \\
&&\left. \left. -\frac 12\left( B_\nu ^{*\delta }H_{\mu \delta }^{*}-B_\mu
^{*\delta }H_{\nu \delta }^{*}\right) -\varepsilon _{\mu \nu \rho \lambda
}\psi ^{*\rho }\chi ^{*\lambda }+\frac 12\left( \psi _\nu ^{*\delta }\chi
_{\mu \delta }^{*}-\psi _\mu ^{*\delta }\chi _{\nu \delta }^{*}\right)
\right) \right)   \nonumber  \label{extn4} \\
&&  \label{extn4}
\end{eqnarray}
where, for a $p$-tensor field

\[
\Phi _i^{*}Q\Phi _i=\frac 1{p!}\Phi _i^{*\mu _1..\mu _p}Q\Phi _{i\mu _1..\mu
_p}\;. 
\]
Following \cite{n4sor}, the gauge-fixing term in the Landau gauge is given
by 
\begin{eqnarray}
S_{{\rm gf}}=Q\,\,{\rm Tr}\int d^4x\left( \overline{c}\partial ^\mu A_\mu
\right) \,+\,4g^2\varepsilon ^{+\mu }\varepsilon ^{-\nu }\,{\rm Tr}\,\int
d^4x\varepsilon _{\mu \nu \rho \lambda }\partial ^\rho \overline{c}%
H^{*\lambda }\,,  \label{GFAction}
\end{eqnarray}
where the antighost $\overline{c},$ introduced by shifting the antifield $%
A_\mu ^{*}$ as $A_\mu ^{*}\rightarrow A_\mu ^{*}+\partial _\mu \overline{c}$%
, is required to transform as 
\begin{eqnarray}
Q\overline{c} &=&b\;,  \nonumber  \label{QAntTransf} \\
Qb &=&(\omega ^{+}\varepsilon ^{+\mu }+\omega ^{-}\varepsilon ^{-\mu
})\partial _\mu \overline{c}\;,  \label{QAntTransf}
\end{eqnarray}
where $b$ is the Lagrange multiplier. Finally, the complete gauge-fixed
action $\Sigma $ 
\begin{equation}
\Sigma =S^{{\rm N=4}}+S_{{\rm ext}}+S_{{\rm gf}}\,\,,  \label{TotAction}
\end{equation}
turns out to obey the following Slavnov-Taylor identity

\begin{eqnarray}
{\cal S}(\Sigma )=\left( \omega ^{+}\varepsilon ^{+\mu }\text{ }+\omega
^{-}\varepsilon ^{-\mu }\right) \Delta _\mu ^{{\rm cl}},  \label{n4slav}
\end{eqnarray}
with 
\begin{equation}
{\cal S}\left( \Sigma \right) ={\rm Tr}\int d^4x\left( \frac{\delta \Sigma }{%
\delta \Phi _i^{*}}\frac{\delta \Sigma }{\delta \Phi _i}+b\frac{\delta
\Sigma }{\delta \overline{c}}+\left( (\omega ^{+}\varepsilon ^{+\mu }+\omega
^{-}\varepsilon ^{-\mu })\partial _\mu \overline{c}\right) \frac{\delta
\Sigma }{\delta b}\right) \;,  \label{stop0}
\end{equation}
and 
\begin{eqnarray}
\Delta _\rho ^{{\rm cl}} &=&{\rm Tr}\int d^4x\left( -A_\mu ^{*}\partial
_\rho A^\mu -H^{*\mu }\partial _\rho H_\mu -\frac 12B^{*\mu \nu }\partial
_\rho B_{\mu \nu }-\tau ^{*}\partial _\rho \tau \right.  \nonumber
\label{sfontes} \\
&&-\frac 12H^{*\mu \nu }\partial _\rho H_{\mu \nu }+\frac 12\psi ^{*\mu \nu
}\partial _\rho \psi _{\mu \nu }+\frac 12\chi ^{*\mu \nu }\partial _\rho
\chi _{\mu \nu }+\psi ^{*\mu }\partial _\rho \psi _\mu \;  \nonumber \\
&&\left. \frac {}{}+\chi ^{*\mu }\partial _\rho \chi _\mu +\xi ^{*}\partial
_\rho \xi +\eta ^{*}\partial _\rho \eta -\phi ^{*}\partial _\rho \phi -%
\overline{\phi }^{*}\partial _\rho \overline{\phi }+c^{*}\partial _\rho
c\right) \,.  \nonumber  \label{que} \\
&&  \label{que}
\end{eqnarray}
As before, $\Delta _\rho ^{{\rm cl}}$ is linear in the quantum fields,
representing a classical breaking not affected by the quantum corrections.
The linearized Slavnov-Taylor operator $B_\Sigma $ 
\begin{equation}
B_\Sigma ={\rm Tr}\int d^4x\left( \frac{\delta \Sigma }{\delta \Phi _i^{*}}%
\frac \delta {\delta \Phi _i}+\frac{\delta \Sigma }{\delta \Phi _i}\frac
\delta {\delta \Phi _i^{*}}+b\frac \delta {\delta \bar{c}}+(\omega
^{+}\varepsilon ^{+\mu }+\omega ^{-}\varepsilon ^{-\mu })\partial _\mu \bar{c%
}\frac \delta {\delta b}\right) \;,  \label{LinSlavReduc1}
\end{equation}
is nilpotent modulo a total derivative 
\begin{equation}
B_\Sigma B_\Sigma =(\omega ^{+}\varepsilon ^{+\mu }+\omega ^{-}\varepsilon
^{-\mu })\partial _\mu \,\;.  \label{LinSlavReduc2}
\end{equation}
Repeating the same steps as in $N=2$ SYM, it turns out that the {\it one to
one} relationship $\left( \ref{n2rel}\right) $ generalizes \cite{n4sor} to

\begin{equation}
g\frac{\partial \Sigma }{\partial g}=-\frac{\varepsilon ^{\mu \nu \rho
\sigma }}{96g^2}{\cal W}_\mu {\cal W}_\nu {\cal W}_\rho {\cal W}_\sigma {\rm %
Tr}\int d^4x\left( \omega ^{+2}\,\phi -\omega ^{-2}\,\overline{\phi }%
\;+\omega ^{+}\omega ^{-}\tau \right) ^2+\;B_\Sigma \Xi ^{-1}\;,
\label{n4act}
\end{equation}
for some local polynomial $\Xi ^{-1}$. In the present case the climbing up
operator ${\cal W}_\mu $ is defined as 
\begin{equation}
{\cal W}_\mu =\frac 12\left[ \left( \frac 1{\omega ^{+}}\frac \partial
{\partial \varepsilon ^{+\mu }}+\frac 1{\omega ^{-}}\frac \partial {\partial
\varepsilon ^{-\mu }}\right) ,B_\Sigma \right] \;,  \label{ClimbDesEq}
\end{equation}
and obeys the same relations $\left( \ref{EBAlgebra}\right) .$

Of course, the proof given in \cite{n2sor} can be repeated straightforwardly
to show that the insertion $\left[ {\rm Tr}\left( \omega ^{+2}\,\phi -\omega
^{-2}\,\overline{\phi }\;+\omega ^{+}\omega ^{-}\tau \right) ^2\cdot \Gamma
\right] $ has indeed vanishing anomalous dimension. Therefore, according to
our theorem, the ultraviolet finiteness of $N=4$ to all orders of
perturbation theory follows from the vanishing of the one-loop beta function 
$\beta _g$, which is very well known since long time \cite{n4l}.

\section{Conclusion}

In this work an algebraic criterion for the ultraviolet finiteness has been
presented. The whole framework relies on the analysis of the descent
equations following from the integrated consistency condition for invariant
counterterms. In some cases, these equations allow to put in {\it one to one}
correspondence the quantized action with a gauge invariant local field
polynomial. The vanishing at the quantum level of the anomalous dimension of
this polynomial leads to the finiteness theorem proven in Sect.3, stating
that if the one-loop order coefficient $\beta _g^{(1)}$ vanishes, then $%
\beta _g$ vanishes to all orders. The knowledge of the one-loop order beta
function $\beta _g^{(1)}$ enables us then to establish whether a given model
can be made ultraviolet finite to all orders of perturbation theory. In
general, the vanishing of $\beta _g^{(1)}$ can be achieved by an appropriate
tuning of the various terms contributing to $\beta _g^{(1)}$, amounting to a
suitable choice of the group representations of the field content of the
model.

This result shares great analogy with the Adler-Bardeen nonrenormalization
theorem for the gauge anomaly. As is well known, the requirement of the
vanishing of the one-loop order coefficient of the gauge anomaly results in
fact in a careful choice for the spinor representations, leading to classify
the so called anomaly free representations.

We also point out that the present algebraic set up has allowed to cover the
case in which the beta function $\beta _g$ receives at most one-loop
contributions, as in the $N=2$ SYM.

Finally, it is worth mentioning that although the finiteness theorem has
been discussed for models with a single coupling constant, it can be
generalized to the case when several couplings are present. Of course, the
derivative of the action $\Sigma $ with respect to each coupling will define
a nontrivial element of the integrated cohomology of the linearized
Slavnov-Taylor operator $B_\Sigma $. The beta functions of those couplings
which can be put in correspondence with unrenormalized local polynomials
belonging to the cohomology of $B_\Sigma $ in the lowest level of the
descent equations will obey the finiteness theorem. On the other hand, the
beta functions of couplings related to nonintegrated cohomology classes in
the first level of the descent equations, corresponding to nontrivial
pointwise invariant lagrangians\footnote{%
This is the case, for instance, of the pure Yang-Mills lagrangian $F_{\mu
\nu }(x)F^{\mu \nu }(x)$ which is pointwise invariant under the gauge
transformations.}, are free to receive quantum corrections.

\section*{Acknowledgements}

The Conselho Nacional de Desenvolvimento Cient\'{\i}fico e Tecnol\'{o}gico
CNPq-Brazil, the Funda{\c {c}}{\~{a}}o de Amparo {\`{a}} Pesquisa do Estado
do Rio de Janeiro (Faperj), the SR2-UERJ and the Funda\c {c}\~{a}o
Os\'{o}rio are acknowledged for the financial support.

\end{document}